\documentclass[reprint,preprintnumbers,longbibliography]{revtex4-2}
\usepackage{orcidlink}
\usepackage{graphicx}
\usepackage{dcolumn}
\usepackage{bm}
\usepackage{textgreek}
\usepackage{placeins}
\usepackage{braket}
\usepackage{physics}
\usepackage{overpic}
\usepackage{amssymb}
\usepackage{amsmath}

\begin{document}

\begin{figure}
\vskip -1.cm
\leftline{\includegraphics[width=0.15\textwidth]{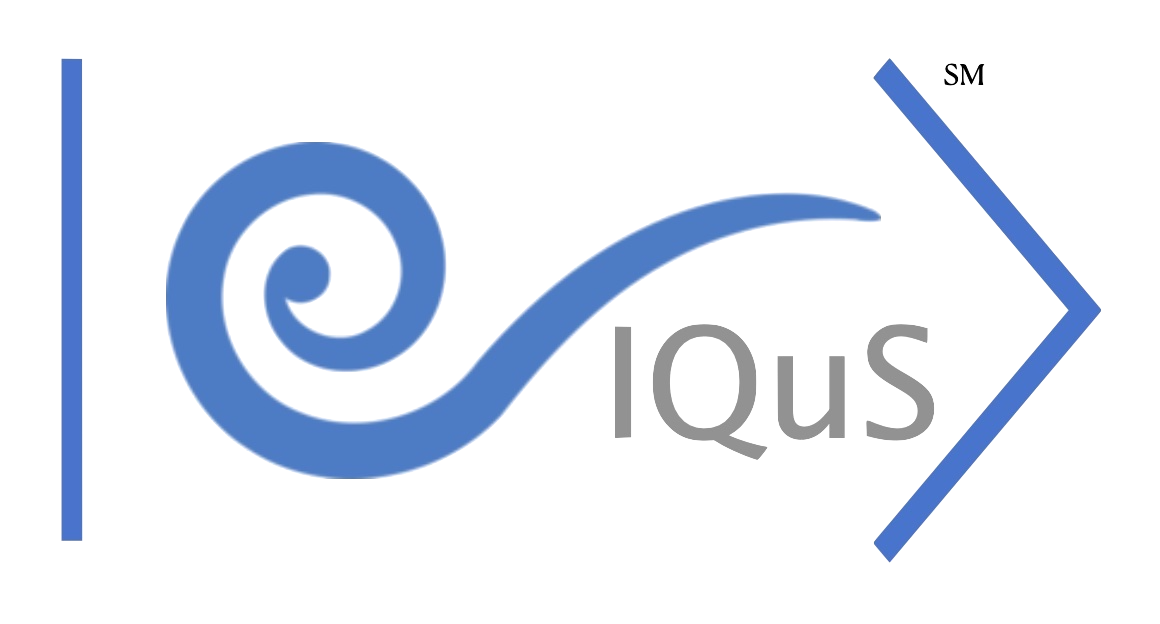}}
\vskip -2.cm
\end{figure}

\preprint{IQuS@UW-21-056}

\title{Quantum Simulation of Lattice QCD with Improved Hamiltonians}

\author{Anthony N. Ciavarella \,\orcidlink{0000-0003-3918-4110}}
 \email{aciavare@uw.edu}
\affiliation{InQubator for Quantum Simulation (IQuS), Department of Physics, University of Washington, Seattle, Washington 98195-1550, USA}%

\date{\today}

\begin{abstract}
Quantum simulations of lattice gauge theories are anticipated to directly probe the real time dynamics of QCD, but scale unfavorably with the required truncation of the gauge fields. Improved Hamiltonians are derived to correct for the effects of gauge field truncations on the SU(3) Kogut-Susskind Hamiltonian. It is shown in $1+1D$ that this enables low chromo-electric field truncations to quantitatively reproduce features of the untruncated theory over a range of couplings and quark masses. In $3+1D$, an improved Hamiltonian is derived for lattice QCD with staggered massless fermions. It is shown in the strong coupling limit that the spectrum qualitatively reproduces aspects of two flavor QCD and simulations of a small system are performed on IBM's {\tt Perth} quantum processor.
\end{abstract}

\maketitle

\section{Introduction}
\noindent
The real time dynamics of quantum chromodynamics (QCD) are of relevance to a number of phenomena in particle and nuclear physics. These range from collisions of hadrons at high energies to the behavior of quark-gluon plasma in the early universe. The simulation of QCD discretized onto a lattice has enabled non-perturbative calculations of static observables in QCD such as hadron masses and form factors~\cite{wilson1974confinement,borsanyi2015ab,karsch2003hadron,tiburzi2017double,beane2015ab,savage2017proton}. 

Quantum computers are expected to be able to directly probe the real time dynamics of quantum field theories. The recent developments in quantum hardware have inspired studies into how to implement simulations of lattice gauge gauge theories on quantum computers. The first quantum simulations of pure non-Abelian lattice gauge theories have been performed in low dimensions on quantum hardware~\cite{klco20202,Rahman:2022rlg,rahman2022real,ciavarella2021trailhead,ciavarella2022preparation,alam2022primitive,Illa:2022jqb,Gustafson_2022}. There have also been quantum simulations of non-Abelian gauge theories coupled to matter in one spatial dimension~\cite{Atas_2021,farrell2023preparations,farrell2023preparations2,Atas:2022dqm}. Theoretical studies have been performed into how to scale up these calculations to larger systems~\cite{byrnes2006simulating,kan2022lattice,shaw2020quantum,stryker2021shearing,davoudi2023general,lamm2019general,zache2023quantum,mathis2020toward,zache2023fermionqudit,gonzalez2022hardware,zohar2015formulation,zohar2013cold,zohar2017digital,bender2018digital,zohar2022quantum,zohar2013quantum,zohar2015quantum,davoudi2021search,halimeh2023spin,halimeh2022gauge,belyansky2023highenergy} and large scale simulations have been performed of Abelian gauge theories~\cite{yang2020observation,Zhou_2022,Su_2023,zhang2023observation,mildenberger2022probing}. However, all these approaches to simulating gauge theories require the gauge field to be truncated and scale poorly with the gauge field truncation. Similar problems were found in the classical simulation of lattice gauge theories with the scaling of errors with lattice spacing. These problems were mitigated through the development of improved Symanzik actions with more favorable scaling of errors with lattice spacing~\cite{Symanzik:1983dc}. Progress has been made towards deriving improved Hamiltonians that reduce errors from finite lattice spacing~\cite{Carena_2022}. It is expected that improved Hamiltonians can be found that mitigate the effects of truncating the gauge field as well.

In this work, improved Hamiltonians are derived for lattice gauge theories through the application of the similarity renormalization group (SRG). SU(3) gauge fields coupled to fermions in 1+1D are used as a case study for the improved Hamiltonians. Tensor network simulations are used to demonstrate that the improved Hamiltonians derived in 1+1D correctly reproduce observables on large lattices. An improved Hamiltonian for lattice QCD with two flavors is derived for 3+1D and a small simulation is performed on IBM's quantum processors.




\section{1+1D}
\subsection{1+1D Hamiltonian}
\noindent
Gauge theories in one spatial dimension have been used as toy models to study the quantum simulation of gauge theories in higher dimensions as they share many qualitative features and their reduced complexity makes simulation more tractable. Previous simulations on quantum hardware have studied the dynamics of hadrons in one spatial dimension~\cite{Atas_2021,farrell2023preparations,Atas:2022dqm} and $\beta$ decay~\cite{farrell2023preparations2}.

In this work, the SU(3) Kogut Susskind Hamiltonian~\cite{kogut1975hamiltonian} with a single flavor of staggered fermions in $1+1D$ will be used as a toy model to study the effects of gauge field truncation and the performance of improved Hamiltonians. The Hamiltonian describing this theory is 
\begin{align}
    & \hat{H} = \hat{H}_{Kin} + \hat{H}_m + \hat{H}_{E} \nonumber \\
    & \hat{H}_{Kin} = \sum_{x,a,b} \frac{1}{2} \hat{\psi}_{x,a}^\dagger \hat{U}^{a,b}_{x,x+1} \hat{\psi}_{x+1,b} + \text{h.c.} \nonumber \\
    & \hat{H}_m = m \sum_{x,a} (-1)^x \hat{\psi}_{x,a}^\dagger \hat{\psi}_{x,a} \nonumber \\
    & \hat{H}_E = \sum_{x,c} \frac{g^2}{2} \hat{E}_{x,x+1}^c\hat{E}_{x,x+1}^c \ \ \ ,
    \label{eq:QCD1DGauge}
\end{align}
where $g$ is the gauge coupling, $m$ is the fermion mass, $\hat{\psi}_{x,a}$ is the fermion field at site $x$ with color $a$,  $\hat{U}^{a,b}_{x,x+1}$ is the parallel transporter on the link between the sites $x,x+1$ and $\hat{E}_{x,x+1}^c$ is the chromo-electric field operator. By working with open boundary conditions in the axial gauge, and enforcing Gauss's law, the gauge fields in this theory can be completely integrated out yielding the Hamiltonian 
\begin{align}
    & \hat{H} = \hat{H}_{Kin} + \hat{H}_m + \hat{H}_{E} \nonumber \\
    & \hat{H}_{Kin} = \sum_{x,a} \frac{1}{2} \hat{\psi}_{x,a}^\dagger \hat{\psi}_{x+1,a} + \text{h.c.} \nonumber \\
    & \hat{H}_m = m \sum_{x,a} (-1)^x \hat{\psi}_{x,a}^\dagger \hat{\psi}_{x,a} \nonumber \\
    & \hat{H}_E = \sum_{x,c} \frac{g^2}{2} \left(\sum_{y<x} \hat{Q}_y^c\right) \left(\sum_{y<x} \hat{Q}_y^c\right)\ \ \ ,
    \label{eq:QCD1DNoGauge}
\end{align}
where $\hat{Q}_y^c$ is the chromo-electric charge at site $x$ defined by
\begin{equation}
    \hat{Q}_y^c = \sum_{a,b} \hat{\psi}^\dagger_{y,a} T^c_{a,b} \hat{\psi}_{y,b} \ \ \ ,
\end{equation}
where $T^c_{a,b}$ are the Gell-Mann matrices. By working with this Hamiltonian, we can directly study the untruncated theory and the performance of improved Hamiltonians that correct for the gauge field truncation.





\subsection{Strong Coupling Expansion $m=0$}
\noindent
Before the Hamiltonian in Eq.~\eqref{eq:QCD1DGauge} can be mapped onto a quantum computer, it must first be truncated to a finite Hilbert space. Typically, this is done by working in the basis of the chromo-electric field and truncating the field below some cutoff. It has been shown numerically for some small systems~\cite{klco2019digitization,ciavarella2021trailhead,zache2023quantum,hayata2023breaking,davoudi2021search} and rigorously proven in general~\cite{tong2022provably} that the error induced by this truncation falls off exponentially with the truncation. The error due to gauge field truncation can be reduced even further by first performing a unitary rotation on the Hamiltonian to reduce the coupling to the higher electric field states and then truncating. In other words, there is a low-energy subspace coupled to a high-energy subspace and one would like to derive an effective field theory description of the low-energy subspace with the high-energy subspace decoupled. Previous work has explored how to perform this decoupling variationally~\cite{wurtz2020variational,robin2023quantum}. One alternative method to construct such an effective Hamiltonian is Schrieffer-Wolff perturbation theory which systematically constructs approximate unitary transformations that decouple the high-energy subspace~\cite{schrieffer1966relation,bravyi2011schrieffer}.

As an example, we will consider the Hamiltonian in Eq.~\eqref{eq:QCD1DGauge} on two staggered sites (one physical site) with massless fermions, truncated at zero electric field. This is the harshest possible truncation that can be applied, and the only physical states left in the Hilbert space are those where sites are unoccupied or have three fermions present forming a color singlet, i.e., a baryon. At this truncation, the Hamiltonian in Eq.~\eqref{eq:QCD1DGauge} is trivial, and there are no dynamics. The states kept in this truncation span the zero electric energy subspace while all states with higher electric energy are being discarded. Using the Schrieffer-Wolff perturbation theory, an effective Hamiltonian for the zero electric energy subspace at leading order is given by
\begin{align}
    \hat{H}_{eff} & = \sum_{x} \frac{9}{16g^2} \hat{Z}_{x} \hat{Z}_{x+1} + \frac{27}{32g^4}\left(\hat{X}_x \hat{X}_{x+1} + \hat{Y}_x \hat{Y}_{x+1}\right) \nonumber \\
    & + \mathcal{O}(g^{-6}) \ \ \ ,
    \label{eq:HeffSW}
\end{align}
where $\hat{X}_x$, $\hat{Y}_x$, $\hat{Z}_x$ are the corresponding Pauli matrices at site $x$ on the lattice. In this basis, spin up states correspond to a site being unoccupied and spin down states correspond to a baryon being present on the site. The details of this derivation and how to systematically derive higher order terms are in Appendix \ref{appendix:SWPT}. In this context, the Schrieffer-Wolff expansion corresponds to performing a strong coupling expansion around the zero electric energy subspace. Note that similar results have been derived for SU(2) lattice gauge theories and the Schwinger model with multiple flavors, showing that they are equivalent to spin systems in the strong coupling limit~\cite{PhysRevB.38.745,berruto1999correspondence,berruto1999quantum}.

\begin{figure}
    \centering
    \includegraphics[width=8.6cm]{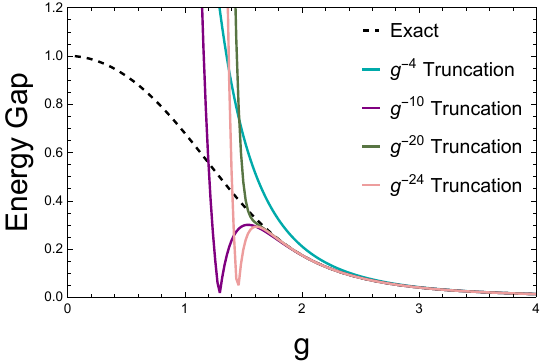}
    \caption{Energy gaps as a function of coupling $g$ for the improved Hamiltonian derived with Schrieffer-Wolff perturbation theory. The black dashed curve is the energy gap of the exact Hamiltonian in Eq.~\eqref{eq:QCD1DNoGauge} and the blue curve is the energy gap of the Hamiltonian in Eq.~\eqref{eq:HeffSW}. The other curves correspond to including higher order terms in the Schrieffer-Wolff expansion of the improved Hamiltonian.}
    \label{fig:SWPT}
\end{figure}

The effective Hamiltonian in Eq.~\eqref{eq:HeffSW} requires only a single qubit per site to be mapped onto a quantum computer. The Hamiltonian in Eq.~\eqref{eq:QCD1DNoGauge} with gauge fields integrated out requires three qubits per site to represent the state of the system. By using this effective Hamiltonian to describe a subspace of the system, the computational resources required are reduced. However, the Schrieffer-Wolff expansion is known to have a finite radius of convergence~\cite{bravyi2011schrieffer}, so this effective Hamiltonian should only be valid over a limited range of couplings. The energy gap for the effective Hamiltonians obtained at different orders in the Schrieffer-Wolff expansion over a range of couplings are shown in Fig.~\ref{fig:SWPT}. Note that both the ground state and first excited state are in the baryon number zero sector. As this figure shows, the effective Hamiltonians obtained through the Schrieffer-Wolff expansion are only valid for strong couplings, and the expansion fails to converge at weak couplings.

\FloatBarrier

\subsection{Similarity Renormalization Group $m=0$}
\noindent
The strong coupling expansion in the previous section was able to yield an improved Hamiltonian to correct for the chromo-electric field truncation for a small system. However, the performance of the improved Hamiltonian was limited by the convergence of the strong coupling expansion. An alternative approach to derive an improved Hamiltonian is the similarity renormalization group (SRG). This method works by choosing a generator of unitary rotations that should decouple the high energy subspace and then continuously flowing to decouple the high energy subspace. The similarity renormalization group was originally introduced to help with the renormalization of light-front quantum field theories~\cite{glazek1997renormalization,WEGNER2000141,szpigel2000similarity}. Similar to the calculations in the previous section, it was found that perturbative calculations in light front field theories suffered from convergence issues and it was found that the SRG could mitigate these issues. It has also been used in low energy nuclear physics to derive low energy nuclear potentials with improved convergence properties~\cite{PhysRevC.75.061001}. Explicitly the Hamiltonian being flowed is parametrized as
\begin{equation}
    \hat{H}_s = \hat{H}_{\Lambda} + \hat{V}_s \ \ \ ,
\end{equation}
where $\hat{H}_{\Lambda}$ determines the energy scales that should be decoupled, $\hat{V}_s$ is the remaining terms in the Hamiltonian and $s$ is the flow parameter. The generator of the SRG flow is traditionally taken to be
\begin{equation}
    \hat{\eta}_s = [\hat{H}_{\Lambda},\hat{H}_s] \ \ \ .
\end{equation}
Note that because $\hat{\eta}_s$ is anti-Hermitian, the evolution under the SRG is unitary and prior to truncation $\hat{H}_s$ is the same as the original Hamiltonian just expressed in a different basis. The evolution of the Hamiltonian under SRG is given by
\begin{align}
        \frac{d\hat{H}_s}{ds} & = \frac{d\hat{V}_s}{ds} = [[\hat{H}_{\Lambda},\hat{V}_s],\hat{H_s}] \nonumber \\
        & = [[\hat{H}_\Lambda,\hat{V}_s], \hat{H}_{\Lambda}] + [[\hat{H}_{\Lambda},\hat{V}_s],\hat{V}_s] \ \ \ .
\end{align}
By flowing to $s\rightarrow\infty$, the low and high energy sectors will be decoupled. It can be seen that these sectors decouple because the commutator $[\hat{H}_\Lambda,\hat{V}_s]$ must vanish as $s\rightarrow\infty$ for the SRG to converge and when this commutator vanishes the two sectors have decoupled. In the following sections, it will be shown how the SRG can be used to derive improved Hamiltonians that correct for the effects of gauge field truncation.

\subsubsection{Two Staggered Sites}
\begin{figure}
    \centering
    \includegraphics[width=8.6cm]{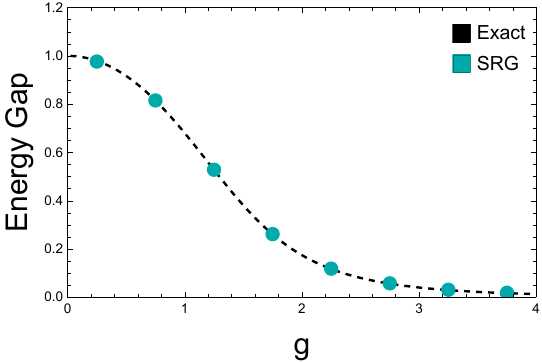}
    \caption{Energy gaps as a function of coupling $g$ for the improved Hamiltonian derived with the SRG. The black dashed curve is the energy gap of the exact Hamiltonian in Eq.~\eqref{eq:QCD1DNoGauge} and the blue points are the energy gap of the Hamiltonian in Eq.~\eqref{eq:HeffSRG2}.}
    \label{fig:SRG2Sites}
\end{figure}

Once again, the Hamiltonian in Eq.~\eqref{eq:QCD1DGauge} on two staggered sites (one physical site), truncated at zero electric field will be used as an example to construct an improved Hamiltonian. The generator of the SRG flow will be chosen to decouple states with different electric energies, i.e. $\hat{H}_{\Lambda} = \hat{H}_E$. The SRG equations can then be solved numerically to recover an improved Hamiltonian of the form
\begin{align}
    \hat{H}_{SRG} & =  A(g)\left(\hat{X}_1 \hat{X}_{2} + \hat{Y}_1 \hat{Y}_{2}\right) + B(g) \hat{Z}_{1} \hat{Z}_{2} \ \ \ ,
    \label{eq:HeffSRG2}
\end{align}
where $A(g)$ and $B(g)$ are constants computed numerically. It can be seen that the improved Hamiltonian must take this form due to the symmetries of the full Hamiltonian. The Hamiltonian in Eq.~\ref{eq:QCD1DGauge} conserves quark number and this symmetry commutes with the truncation of the Hamiltonian. Combined with the chiral symmetry of this Hamiltonian at $m=0$, the only allowed operators in the improved Hamiltonian are $\hat{X}_1 \hat{X}_2 + \hat{Y}_1 \hat{Y}_2$ and $\hat{Z}_1 \hat{Z}_2$. Note that this Hamiltonian takes the same form as that derived in the strong coupling expansion in Eq.~\eqref{eq:HeffSW} except now the coefficients multiplying the operators have been determined through SRG instead of a perturbative expansion. The energy gap for this Hamiltonian as a function of the coupling is shown in Fig.~\ref{fig:SRG2Sites}. Unlike the improved Hamiltonian obtained through the strong coupling expansion, the improved Hamiltonian obtained through the SRG suffers from no convergence issues and is able to correctly reproduce the energy gap at all values of the coupling.

\subsubsection{Larger Systems}
As shown in the previous section, the SRG was capable of producing an improved Hamiltonian that correctly describes the physics of a small system. In practice, improved Hamiltonians will be needed for larger systems. The setup of the SRG used in the previous section does not scale efficiently to larger lattices. This is because as the SRG evolves, the number of operators generated can be exponential in the system size. This can be mitigated through the use of the in-medium similarity renormalization group (IMSRG) which truncates operators in the SRG flow above a certain weight~\cite{hergert2016medium}. In practice IMSRG can be implemented by performing SRG on a lattice of the same size as the operator size truncation and matching the obtained improved Hamiltonian to a translationally invariant Hamiltonian defined on a larger lattice. The cost of performing the IMSRG scales exponentially with the size truncation. However the convergence with operator size is also exponential due to the exponential decay of correlations in low energy states. 

As an explicit example, improved Hamiltonians for the zero electric field truncation will be derived with IMSRG. The smallest nontrivial operator size truncation is at two staggered sites. The improved Hamiltonian derived with IMSRG at this truncation with coupling $g$ on $L$ staggered sites is
\begin{align}
    \hat{H}_{SRG} = & \sum_{x < L} A(g)\left(\hat{X}_x \hat{X}_{x+1} + \hat{Y}_x \hat{Y}_{x+1}\right)  \nonumber \\ 
    & + B(g) \hat{Z}_{x} \hat{Z}_{x+1} \ \ \ .
\end{align}
This improved Hamiltonian comes from doing the same calculation as the previous section and matching the two staggered site improved Hamiltonian to a translationally invariant one defined on a larger lattice. The accuracy of the improved Hamiltonians derived through IMSRG at this electric field truncation can be improved by computing the IMSRG flow for larger operator size truncations. In general, one would expect this method to work well when the operator size truncation used is comparable to the correlation length of the system in question. Explicitly, the form of the improved Hamiltonians obtained by truncating at operators defined on three staggered sites takes the form
\begin{align}
    \hat{H}_{3,SRG} = & \sum_x A_1(g) \left(\hat{X}_x \hat{X}_{x+1} + \hat{Y}_x \hat{Y}_{x+1}\right) + B_1(g) \hat{Z}_{x} \hat{Z}_{x+1} \nonumber \\
    & + A_2(g) \left(\hat{X}_x \hat{X}_{x+2} + \hat{Y}_x \hat{Y}_{x+2}\right) + B_2(g) \hat{Z}_{x} \hat{Z}_{x+2}
    \label{eq:SRG3}
\end{align}
where $A_i(g)$, and $B_i(g)$ are constants determined from solving the SRG equations numerically. Note that this takes the same form as Eq.~\eqref{eq:HeffSRG2} just with the inclusion of next to nearest neighbor hopping. The performance of the improved Hamiltonians can be improved further by truncating the operator size at four staggered sites. The improved Hamiltonian obtained at this truncation takes the form
\begin{align}
    \hat{H}_{4,SRG} = & \sum_x A_1(g) \left(\hat{b}_x \hat{b}^\dagger_{x+1} + \hat{b}^\dagger_x \hat{b}_{x+1}\right) +B_1(g) \hat{Z}_x \hat{Z}_{x+1} \nonumber \\
    & + A_2(g) \left(\hat{b}_x \hat{b}^\dagger_{x+2} + \hat{b}^\dagger_x \hat{b}_{x+2}\right) + B_2(g) \hat{Z}_x \hat{Z}_{x+2} \nonumber \\
    & + A_3(g) \left(\hat{b}_x \hat{b}^\dagger_{x+3} + \hat{b}^\dagger_x \hat{b}_{x+3}\right) + B_3(g) \hat{Z}_x \hat{Z}_{x+3} \nonumber \\
    & + C_1(g) \left(\hat{b}_x \hat{b}^\dagger_{x+1} + \hat{b}^\dagger_x \hat{b}_{x+1}\right) \hat{Z}_{x+2} \hat{Z}_{x+3} \nonumber \\
    & + C_2(g) \left(\hat{b}_x \hat{b}^\dagger_{x+2} + \hat{b}^\dagger_x \hat{b}_{x+1}\right) \hat{Z}_{x+1} \hat{Z}_{x+3} \nonumber \\
    & + C_2(g) \left(\hat{b}_{x+1} \hat{b}^\dagger_{x+3} + \hat{b}^\dagger_{x+1} \hat{b}_{x+3}\right) \hat{Z}_{x} \hat{Z}_{x+2} \nonumber \\
    & + C_3(g) \left(\hat{b}_x \hat{b}^\dagger_{x+3} + \hat{b}^\dagger_x \hat{b}_{x+1}\right) \hat{Z}_{x+1} \hat{Z}_{x+2} \nonumber \\
    & + C_4(g) \left(\hat{b}_{x+1} \hat{b}^\dagger_{x+2} + \hat{b}^\dagger_{x+1} \hat{b}_{x+2}\right) \hat{Z}_{x} \hat{Z}_{x+3} \nonumber \\
    & + C_5(g) \hat{Z}_{x} \hat{Z}_{x+1} \hat{Z}_{x+2} \hat{Z}_{x+3} \nonumber \\
    & + D_1(g) \left(b^\dagger_{x} b^\dagger_{x+1} b_{x+2} b_{x+3} + b_{x} b_{x+1} b^\dagger_{x+2} b^\dagger_{x+3} \right) \nonumber \\
   & + D_2(g) \left(b^\dagger_{x} b_{x+1} b_{x+2}^\dagger b_{x+3} + b_{x} b^\dagger_{x+1} b_{x+2} b^\dagger_{x+3} \right) \nonumber \\
   & + D_3(g) \left(b^\dagger_{x} b_{x+1} b_{x+2} b_{x+3}^\dagger + b_{x} b^\dagger_{x+1} b^\dagger_{x+2} b_{x+3} \right)
   \label{eq:SRG4}
\end{align}
where $\hat{b}_x = \frac{1}{2}\left(\hat{X}_x + i \hat{Y}_x\right)$ is a qubit annihilation operator at site $x$ and $A_i(g)$, $B_i(g)$, $C_i(g)$, and $D_i(g)$ are constants determined from solving the SRG equations numerically. Note that it can be seen that the improved Hamiltonians must take the forms in Eq.~\ref{eq:SRG3} and Eq~\ref{eq:SRG4} at these size truncations by considering the symmetries of the full Hamiltonian as discussed in the previous section.

\begin{figure}
    \centering
    \includegraphics[width=8.6cm]{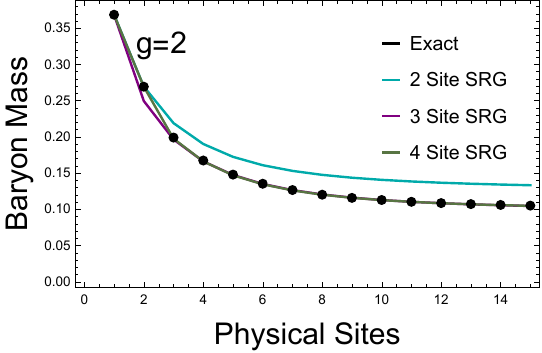}
    \includegraphics[width=8.6cm]{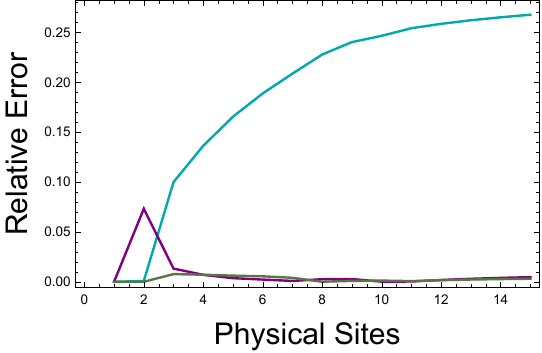}
    \caption{Baryon mass as a function of lattice size for $g=2$. The black points show the baryon mass for the Hamiltonian in Eq.~\eqref{eq:QCD1DNoGauge}. The different solid curves correspond to the baryon mass in the various improved Hamiltonians derived through the use of IMSRG with different operator size truncations.}
    \label{fig:SRGLargeG2}
\end{figure}

\begin{figure}
    \centering
    \includegraphics[width=8.6cm]{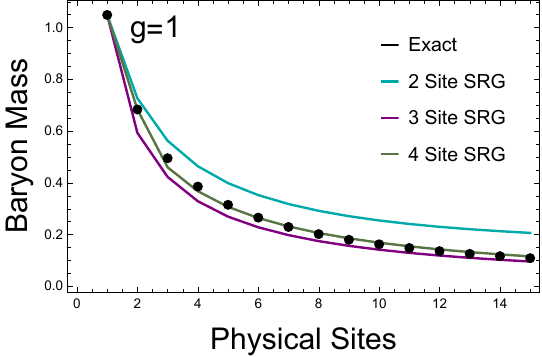}
    \includegraphics[width=8.6cm]{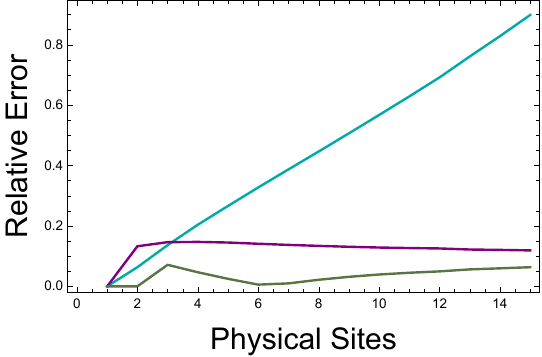}
    \caption{Baryon mass as a function of lattice size for $g=1$. The black points show the baryon mass for the Hamiltonian in Eq.~\eqref{eq:QCD1DNoGauge}. The different solid curves correspond to the baryon mass in the various improved Hamiltonians derived through the use of IMSRG with different operator size truncations.}
    \label{fig:SRGLargeG1}
\end{figure}

To test the performance of the improved Hamiltonians derived through SRG, density matrix renormalization group (DMRG) calculations were performed using the C++ {\tt iTensor} library~\cite{itensor-r0.3,fishman2022itensor,white1992density,white1993density,verstraete2004matrix} to obtain the vacuum state and the single baryon ground state of the Hamiltonian in Eq.~\eqref{eq:QCD1DNoGauge} and the improved Hamiltonians described above for lattices with up to fifteen physical sites with open boundary conditions. Fig.~\ref{fig:SRGLargeG2} shows the mass of the baryon (difference of the energy of the single baryon state and vacuum state) for the full Hamiltonian and the improved Hamiltonians for the zero electric field truncation for $g=2$. As this figure shows, the relative error in the baryon mass computed with the improved Hamiltonians grows with system size and then saturates. By using improved Hamiltonians with a larger operator size truncation in the IMSRG, the relative error in the baryon mass can be reduced down to the percent level. The baryon mass for $g=1$ was also computed and is shown in Fig.~\ref{fig:SRGLargeG1}. At this weaker coupling, the correlation length is longer and the relative error in the baryon mass grows uncontrollably with the lattice size for the improved Hamiltonian obtained by the two staggered site truncation IMSRG. However, increasing the size of the operator truncation used in the IMSRG decreases the error in the baryon mass to controllable levels. This rapid convergence with the operator size is due to the IMSRG correctly reproducing all correlations smaller than the operator size truncation. It is expected that once the operator size truncation is larger than the inverse baryon mass, the improved Hamiltonian derived through IMSRG will accurately describe single baryon states. In general, it is expected that the improved Hamiltonians derived with IMSRG will accurately describe states containing correlations that are smaller than the operator size truncation.

\begin{figure}
    \centering
    \includegraphics[width=8.6cm]{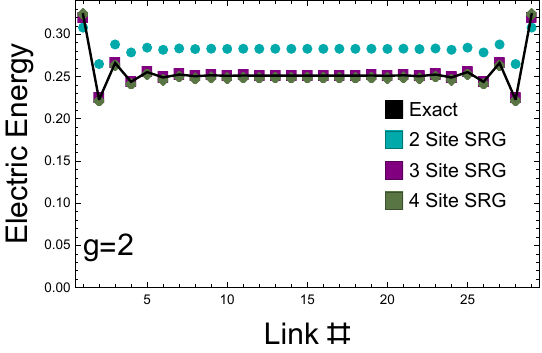}
    \includegraphics[width=8.6cm]{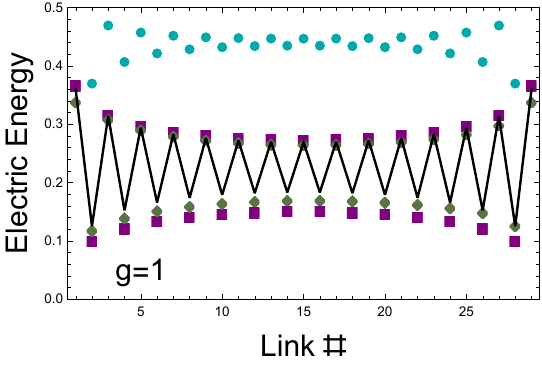}
    \caption{The expectation of the electric energy on each link for a lattice with 15 physical sites. The black points were computed using the Hamiltonian in Eq.~\eqref{eq:QCD1DNoGauge}. The other points were computed using the improved Hamiltonians for the zero electric field truncation computed using SRG.}
    \label{fig:ElectricSRG}
\end{figure}

In addition to studying the energy of different states on the lattice, the IMSRG flows of operators can be computed and their expectation values can be computed using improved Hamiltonians. As an explicit example, the SRG flow of the chromo-electric energy density was computed. The operators corresponding to the chromo-electric operators in the improved basis are the same as those that show up in the improved Hamiltonians, just with different coefficients. The vacuum expectation of the chromo-electric energy density is shown in Fig.~\ref{fig:ElectricSRG} for $g=1$ and $g=2$. As before, increasing the size of the operator truncation in the IMSRG improves the accuracy of the improved Hamiltonians. Remarkably, even though the improved Hamiltonians are being truncated at zero electric field, their ground states still reproduce the electric energy density of the full untruncated theory.

\subsection{Similarity Renormalization Group $m \neq 0$}
In the previous section, IMSRG was used to derive an improved Hamiltonian that describes the dynamics of baryons in QCD in one dimension with massless quarks. The same technique can be used to setup improved Hamiltonians in the case of massive quarks as well.

In a theory with massive quarks, the piece of the Hamiltonian that should be used to generate the SRG flow is the combination of the mass and electric energy terms. At the zero electric energy truncation, the only state left after truncation is the one with matter sites empty and anti-matter sites filled. Therefore with massive quarks, there are no dynamics at this level of truncation. The next lowest truncation in the SRG flow depends on the relative size of the fermion mass $m$ and the coupling $g$. If $\frac{2}{3}g^2 > m$, then the next lowest lying state in the spectrum consists of a baryon at a site. The improved Hamiltonian derived by truncating at this level takes the same form as in the previous section except with the addition of a mass term for the baryons. If instead $\frac{2}{3}g^2 < m$, then the next lowest lying state in the spectrum corresponds to a quark anti-quark pair connected by a link of electric flux. In the strong coupling limit, this corresponds to a meson at the excited link. 

\begin{figure}
    \centering
    \includegraphics[width=8.6cm]{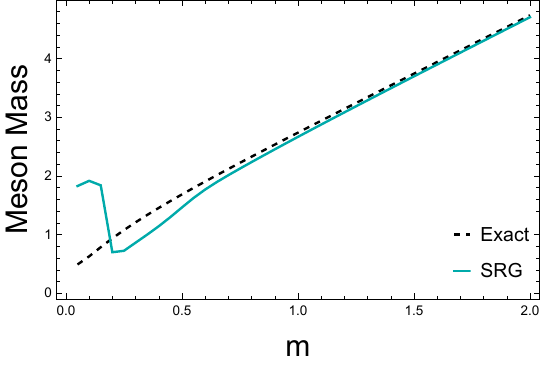}
    \caption{Meson mass as a function of quark mass $m$ for $g=1$ on a lattice of 2 physical sites (4 staggered). The black dashed curves shows the meson mass for the Hamiltonian in Eq.~\eqref{eq:QCD1DNoGauge}. The blue curve shows the meson mass for the improved Hamiltonian in Eq.~\eqref{eq:SRG2Links}.}
    \label{fig:SRGMeson}
\end{figure}

Denoting the trivial vacuum state by $\ket{Vac}$, and the state with a $q\Bar{q}$ pair on link $l$ by $\ket{l}$, the Hamiltonian obtained under IMSRG flow truncating the energy at single link excitations and the operator size at two link operators takes the form
\begin{align}
    & \hat{H}_{SRG} = E_0(g,m) \ket{Vac} \bra{Vac} \nonumber \\
    & + \sum_l h(g,m)\left(\ket{l+1}\bra{l} + \ket{l}\bra{l+1}\right) + E_1(g,m) \ket{l}\bra{l} \ \ \ ,
    \label{eq:SRGpi1}
\end{align}
where $E_0(g,m)$, $E_1(g,m)$, and $h(g,m)$ are constants determined through numerically solving the SRG flow. Note that this Hamiltonian has the same form as that of a single non-relativistic particle. The Hamiltonian in Eq.~\eqref{eq:SRGpi1} can be viewed as a Hamiltonian for a single link excitation (or meson) and can be mapped onto a second quantized Hamiltonian to describe a system with more excited links. Explicitly, the single excitation sector of 
\begin{align}
     \hat{H}_{SRG} & = \sum_l \frac{h(g,m)}{2} \left(\hat{X}_l \hat{X}_{l+1} + \hat{Y}_l \hat{Y}_{l+1} \right) \nonumber \\
    & + \frac{E_0(g,m) - E_1(g,m)}{2} \hat{Z}_l \ \ \ ,
    \label{eq:SRGLinks}
\end{align}
will be identical to the Hamiltonian in Eq.~\eqref{eq:SRGpi1}. This improved Hamiltonian will also be capable of describing states with multiple links excited as well. The description of these states with multiple links excited can be improved by raising the truncation of states kept after SRG flow to include states where two links are excited. By keeping these states after the SRG flow and keeping the other truncations as before, the improved Hamiltonian given by
\begin{align}
    \hat{H}_{SRG2} & = \sum_l \frac{h(g,m)}{2} \left(\hat{X}_l \hat{X}_{l+1} + \hat{Y}_l \hat{Y}_{l+1} \right) \nonumber \\
    & + s(g,m)\hat{Z}_l \hat{Z}_{l+1} + \frac{E_0(g,m) - E_1(g,m)}{2} \hat{Z}_l \ \ \ ,
    \label{eq:SRG2Links} 
\end{align}
will have single and two excitation sectors that match the improved Hamiltonians derived through SRG. As a test of the performance of this improved Hamiltonian, the mass of the meson was computed on a lattice with two physical sites for $g=1$ and various values of $m$ in Fig.~\ref{fig:SRGMeson}. Similar to the massless case, the improved Hamiltonian derived with the SRG performs well when there is a large separation in energy scales between the states being decoupled. Note that in principle, the same comparison can be done with larger lattices, however the meson is in the same baryon number sector as the vacuum which complicates the calculation of the meson mass. It is expected that this improved Hamiltonian scales to larger lattices as in the massless case.

\subsubsection{Quantum Simulation}
As an example of how these improved Hamiltonians can be used for quantum simulation, a simulation will be performed of a meson's time evolution on three physical sites with open boundary conditions. Using the Hamiltonian in Eq.~\eqref{eq:QCD1DNoGauge} would require a quantum computer with 18 qubits to encode the state, and non-local interactions between the qubits to implement the electric energy piece of the Hamiltonian. Using the improved Hamiltonian in Eq.~\eqref{eq:SRG2Links} requires only 5 qubits to represent the state and only requires nearest neighbor interactions on the quantum computer to perform time evolution. 
\begin{figure}
    \centering
    \includegraphics[width=8.6cm]{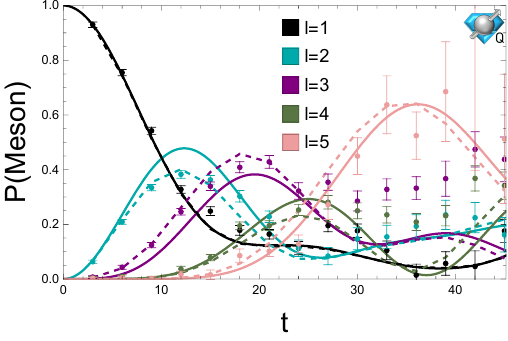}
    \caption{Time evolution of a single meson on 3 physical sites performed on the IBM {\tt Perth} quantum processor. Each color corresponds to the probability of a different link being excited. The solid lines show the exact time evolution. The dashed lines show a classical simulation of the Trotterized time evolution that was implemented on the quantum processor. The data points were obtained using self-mitigating circuits on IBM {\tt Perth}.}
    \label{fig:PerthMeson}
\end{figure}

Fig.~\ref{fig:PerthMeson} shows the real time evolution of a single meson on three physical sites with $g=1,m=1$ simulated on IBM's {\tt Perth} quantum processor~\cite{aleksandrowicz2019qiskit,ibmPerth}. A meson state was prepared on the quantum processor by applying an $\hat{X}$ gate to the qubit assigned to the leftmost link. Time evolution was performed using a first order Trotter formula. Explicitly, the Hamiltonian was decomposed as $\hat{H}=\sum_{l=1}^{4} \hat{H}_l$ where
\begin{equation}
    \hat{H}_l = \frac{h(g,m)}{2} \left(\hat{X}_l \hat{X}_{l+1} + \hat{Y}_l \hat{Y}_{l+1} \right) + s(g,m)\hat{Z}_l \hat{Z}_{l+1} \ \ \ ,
\end{equation}
and the Trotterized time evolution operator was given by
\begin{equation}
    \hat{U}(\Delta t) = e^{-i\hat{H}_2 \Delta t} e^{-i\hat{H}_4 \Delta t} e^{-i\hat{H}_3 \Delta t} e^{-i\hat{H}_1 \Delta t} \ \ \ .
\end{equation}
Each individual $e^{-i\hat{H}_l \Delta t}$ was decomposed into a circuit with 3 CNOT gates using standard techniques~\cite{PhysRevA.69.010301,PhysRevA.77.066301}. The sum over Pauli $\hat{Z}$ operators can be ignored when performing time evolution because it commutes with the full Hamiltonian and the operators being measured. The noise in the quantum simulation was mitigated using self-mitigation combined with Pauli twirling~\cite{urbanek2021mitigating,rahman2022real,Rahman:2022rlg}. For each Trotter step, $50$ circuits describing the time evolution were used along with $50$ circuits with $\Delta t = 0$ used to determine the strength of the depolarizing noise channel. Each circuit was sampled $10,000$ times. As Fig.~\ref{fig:PerthMeson} shows, the quantum hardware is able to describe the time evolution well at short times, but at long times the hardware noise begins to dominate. However, despite the presence of hardware noise at late times, the location of the peak of the wavepacket of the meson can still be located at late times.

\section{3+1D}
\subsection{3+1D Hamiltonian}
\noindent
Performing a quantum simulation of lattice QCD requires a choice of Hamiltonian to be used. This choice is complicated by the phenomena of fermion doubling, where the naive discretization of the Dirac field on the lattice in $d$ dimensions actually describes $2^d$ fermions. Furthermore, the Nielson-Ninomiya theorem forbids the presence of chiral symmetry on the lattice when all doublers are removed~\cite{nielsen1981no,nielsen1981absence}. In this work, staggered fermions will be used. Staggered fermions work by distributing the components of the Dirac field across different sites of the lattice. This preserves some chiral symmetry at the cost of still having some fermion doublers remain. In lattice QCD calculations on classical computers, space and time are both discretized leading to staggered fermions describing 4 types of fermions, referred to as tastes in the literature. For practical calculations, these can be reduced to a single flavor through the process of rooting~\cite{sharpe2006rooted,golterman2008qcd,Bazavov_2013,Follana_2007,RootValidity}. In quantum simulation, time is left continuous and only space is discretized. This changes the counting of the number of tastes present. Explicitly, with three dimensions of space discretized and time left continuous, staggered fermions describe two tastes. This is a feature, not a bug for using lattice QCD to study nuclear physics as one taste can be identified as an up quark and the other can be identified as a down quark. Therefore, we would expect lattice QCD with a single staggered fermion on a quantum computer to describe two flavor QCD where both quarks have the same mass. With massless quarks, this lattice regularization should reproduce the predictions of chiral perturbation theory as the continuum limit is approached. Explicitly, the Hamiltonian that should be used for $3+1$ dimensional two flavor massless lattice QCD on a quantum computer is
\begin{align}
    \hat{H} &= \hat{H}_K + \hat{H}_E + \hat{H}_B \nonumber \\
    \hat{H}_K &= \sum_{\Vec{r},\hat{\mu},a,b} \eta_{\Vec{r},\hat{\mu}}\frac{1}{2} \hat{\psi}_{\Vec{r},a}^\dagger \hat{U}^{a,b}_{\Vec{r},\Vec{r}+\hat{\mu}} \hat{\psi}_{\Vec{r}+\hat{\mu},b}+\text{h.c.} \nonumber \\
    \hat{H}_E &= \frac{g^2}{2}\sum_{l \in \text{links},c}  \hat{E}_{l}^c \hat{E}_{l}^c \nonumber \\
    \hat{H}_B &= -\frac{1}{2g^2} \sum_{p \in \text{plaquettes}} \Box_p \ \ \ ,
    \label{eq:LQCD3}
\end{align}
where $\psi_{\Vec{r},a}$ is a fermion field at site $\Vec{r}$ with color $a$, $\hat{\mu}$ is a unit vector in the $\hat{x}$, $\hat{y}$, or $\hat{z}$ directions, $\eta_{\Vec{r},\hat{\mu}}$ are the spin diagonalization phases, $\hat{U}^{a,b}_{\Vec{r},\Vec{r}+\hat{\mu}}$ is an SU(3) parallel transporter between sites $\Vec{r}$ and $\Vec{r}+\hat{\mu}$, $\hat{E}_{l}^c$ is the SU(3) chromo-electric field on link $l$ and $\Box_p$ is the Hermitian component of the trace over color indices of the product of parallel transporters on plaquette $p$. Previous work has shown that this Hamiltonian has a discrete chiral symmetry corresponding to translation by one lattice site that is spontaneously broken and an isospin symmetry that corresponds to diagonal translations~\cite{banks1977strong,jones1979lattice,schreiber1994t}.
\subsection{Improved Hamiltonian}
As is the case for 1D QCD, mapping the Hamiltonian in Eq.~\eqref{eq:LQCD3} onto qubits is challenging, especially if one wishes to perform a quantum simulation with existing hardware. Improved Hamiltonians can also be derived for performing quantum simulations of this theory. Following the discussions of the previous sections, IMSRG can be applied to this theory with a truncation in operator size. The smallest non-trivial operator size IMSRG can be applied to is a single link and the lowest electric field truncation that can be used is zero electric field. The resulting improved Hamiltonian on the 3 dimensional lattice will take the same form as in the 1D case except now the hopping terms will have phases that result from the spin diagonalization.

Explicitly, the improved Hamiltonian obtained through SRG at this truncation in operator size and electric field is
\begin{align}
    \hat{H}_{SRG} & = \sum_{\Vec{r}} A(g) \left(\hat{\psi}_{\Vec{r}}^\dagger \hat{\psi}_{\Vec{r}+\hat{x}} + \hat{\psi}_{\Vec{r}+\hat{x}}^\dagger \hat{\psi}_{\Vec{r}} \right) \nonumber \\
    & + A(g) (-1)^{r_1}\left(\psi_{\Vec{r}}^\dagger \hat{\psi}_{\Vec{r}+\hat{y}} + \hat{\psi}_{\Vec{r}+\hat{y}}^\dagger \hat{\psi}_{\Vec{r}} \right) \nonumber \\
    & + A(g) (-1)^{r_1 + r_2}\left(\hat{\psi}_{\Vec{r}}^\dagger \hat{\psi}_{\Vec{r}+\hat{z}} + \hat{\psi}_{\Vec{r}+\hat{z}}^\dagger \hat{\psi}_{\Vec{r}} \right) \nonumber \\
    & + B(g) \sum_{\hat{\mu}} \left(2\hat{\psi}^\dagger_{\Vec{r}} \hat{\psi}_{\Vec{r}}-1\right) \left(2\hat{\psi}^\dagger_{\Vec{r}+\hat{\mu}} \hat{\psi}_{\Vec{r}+\hat{\mu}}-1\right) \ \ \ ,
    \label{eq:effQCD}
\end{align}
where $\psi_{\Vec{r}}$ is a colorless fermion field at site $\Vec{r}$ and $A(g)$ and $B(g)$ are numerical constants determined through solving the SRG equations.

Note that this improved Hamiltonian only describes the QCD Hamiltonian accurately for large coupling $g$ as the lattice size is increased. This is because as $g$ is decreased, the correlation length increases. To accurately describe the QCD Hamiltonian at these couplings without increasing the electric field truncation, higher weight operators must be included as in the $1$D case. Alternatively, the electric field truncation could be increased. At large coupling, the $\pi$ meson is massive and is integrated out of this improved Hamiltonian. By increasing the chromo-electric field truncation of states kept after the SRG flow, states with quark-antiquark pairs separated by a link will be included in the low energy Hilbert space kept after truncation and will yield an improved Hamiltonian that describes meson degrees of freedom as well.

\subsubsection{Spectrum}
\noindent
The improved Hamiltonian in Eq.~\eqref{eq:effQCD} will describe the untruncated theory accurately in the limit of large $g$. While the continuum limit of lattice QCD is in the limit of $g\rightarrow0$, large couplings can be used to study the theory at finite lattice spacing. In the limit $g\rightarrow \infty$, $A(g)\rightarrow 0$ and some qualitative features of low energy QCD are recovered. In particular, it has been shown that in the strong coupling limit this theory has an isospin symmetry and a spontaneously broken chiral symmetry~\cite{banks1977strong,jones1979lattice,schreiber1994t}. In addition to the previously studied features of this regularization, the strong coupling limit of this Hamiltonian also reproduces the approximate Wigner SU(4) spin flavor symmetry of nuclear physics.

As an example, we will study the improved Hamiltonian in Eq.~\eqref{eq:effQCD} on a single cube. The fermionic fields will be mapped onto qubits using a Jordan-Wigner encoding. When $A(g)=0$, the Hamiltonian in Eq.~\eqref{eq:effQCD} can be rewritten in terms of Pauli matrices as
\begin{equation}
    \hat{H}_{SRG} = \frac{9}{16g^2} \sum_{\hat{\mu}} \hat{Z}_{\Vec{r}} \hat{Z}_{\Vec{r}+\hat{\mu}} \ \ \ .
\end{equation}
The ground state is in the baryon number $B=0$ sector and is a degenerate N\'{e}el state. For the rest of this discussion, we will only consider the sector that is even under reflection across the $\hat{z}$ axis. The lowest lying excited states in the $B=0$ sector correspond to performing a SWAP operation on one of the links. Denoting the energy cost of flipping one link as $\Delta=\frac{9}{8g^2}$, this set of excited states has energy $4\Delta$ and there are 12 of them. These 12 states should correspond to spin one and spin zero baryon anti-baryon pairs, i.e., $p\Bar{p}$, $n\Bar{n}$, $n\Bar{p}$ and $p\Bar{n}$ states.

The lowest lying energy states in the $B=1$ sector correspond to flipping one site from the N\'{e}el state on the cube. There are four corners that can be flipped in the N\'{e}el state to end up in the $B=1$ sector so there are four degenerate states with energy $3\Delta$. These correspond to the two spin modes of the proton and neutron. Note that the proton and neutron mass are degenerate which should be expected from isospin symmetry. 

In the $B=2$ sector, the lowest lying states correspond to flipping two spins in the N\'{e}el state. This results in six degenerate states with energy $6\Delta$. These states correspond to spin 1 $pn$ states and spin $0$ $pp$, $pn$ and $nn$ states. The fact that these states are degenerate is reflective of spin-flavor symmetry which is approximately present in low energy nuclear physics. The spin-flavor symmetry has been shown to emerge in the large $N_c$ limit of QCD~\cite{KAPLAN1996244} and is related to the minimization of entanglement in low energy nucleon scattering~\cite{beane2019entanglement,Beane:2021zvo,Beane:2020wjl,Bai:2022hfv,Liu:2022grf,Low:2021ufv}. We also see that in the strong coupling limit, the deuteron has binding energy zero. Similar calculations can be done in the higher baryon number sectors which also show that these sectors also demonstrate spin-flavor symmetry and nuclei with binding energy = 0. It is also interesting to note that the nucleon-nucleon scattering lengths are large. As a result, the pionless EFT describing nucleon scattering is an expansion around a non-trivial fixed point where the binding energy of nuclei vanishes as is the case in this lattice regularization~\cite{kaplan1998new,kaplan1998two,van1999effective,birse1999renormalisation}.

\subsubsection{Quantum Simulation}
\noindent
The Hilbert space describing the Hamiltonian in Eq.~\eqref{eq:effQCD} consists of a single fermion mode for each site. Using the Jordan-Wigner encoding, the state of each site can be represented with a single qubit. In this encoding, a list of fermion operators $\psi_1,\hat{\psi}_2,...,\hat{\psi}_{N}$ are mapped onto qubit operators as 
\begin{equation}
 \hat{\psi}_n = \bigotimes_{k<n} \frac{1}{2} \hat{Z}_k \left(\hat{X}_n + i \hat{Y}_n\right) \ \ \ .  
\end{equation}
For a local one dimensional fermionic theory, this fermion encoding leads to a Hamiltonian that is local in qubits. However, in higher dimensions, the operators in the Hamiltonian will include strings of Pauli $\hat{Z}$ operators that wrap around the lattice. These long range operators are necessary to enforce the anti-commutation relations of the fermionic operators and may make it difficult to practically scale to calculations on a large lattice. 

\begin{figure}
    \centering
    \includegraphics[width=8.6cm]{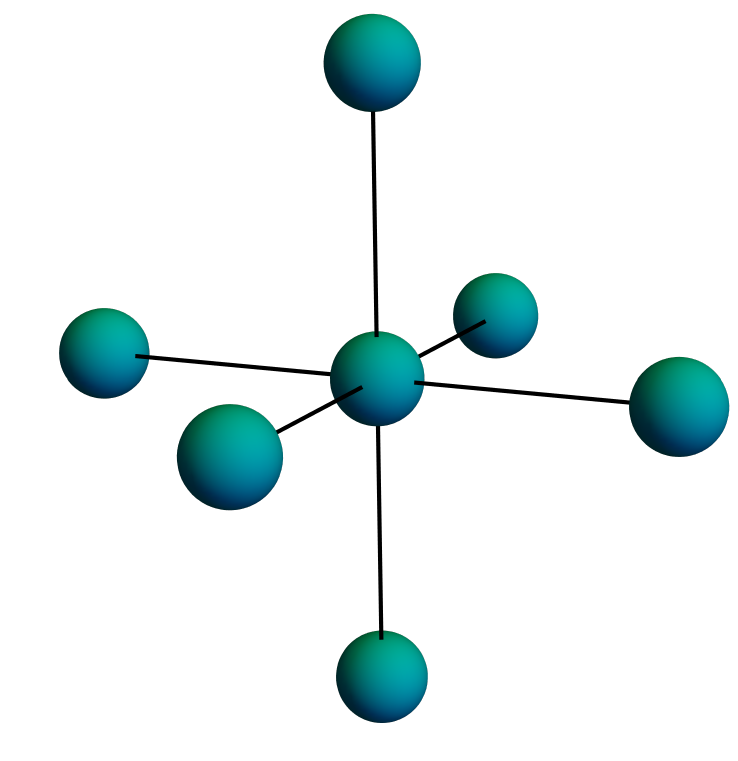}
    \caption{Connectivity of the system described by the improved Hamiltonian in Eq.~\eqref{eq:effQCDV}.}
    \label{fig:Vertex3D}
\end{figure}

\begin{figure}
    \centering
    \includegraphics[width=8.6cm]{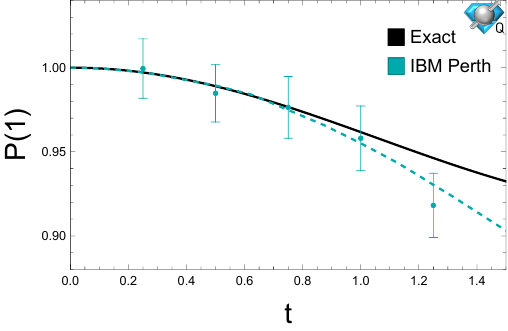}
    \caption{Probability of staying in the trivial vacuum state computed on the IBM {\tt Perth} quantum processor. The solid black line shows the exact time evolution. The blue dashed line shows the probability computed using a first order Trotter step computed on a classical computer. The blue data points were obtained using self-mitigating circuits on IBM {\tt Perth}.}
    \label{fig:Perth3D}
\end{figure}

As a demonstration of how this improved Hamiltonian works in practice, time evolution on six vertices connected to a single vertex at the center as shown in Fig.~\ref{fig:Vertex3D}  will be simulated. This is the smallest non-trivial subsystem of a full three dimensional lattice that will be repeated periodically and will be useful for understanding how simulations on a larger lattice will work. Each of the seven vertices can be mapped onto a single qubit. The Hamiltonian describing their time evolution is given by
\begin{align}
    \hat{H}_{SRG} & = \sum_{v} A(g) \left(\hat{\psi}_{0}^\dagger \hat{\psi}_{v} + \hat{\psi}_{v}^\dagger \hat{\psi}_{0} \right) \nonumber \\
    & + B(g) \sum_{v} \left(2\hat{\psi}^\dagger_{v} \hat{\psi}_{v}-1\right) \left(2\hat{\psi}^\dagger_{0} \hat{\psi}_{0}-1\right) \ \ \ ,
    \label{eq:effQCDV}
\end{align}
where the $0$ subscript denotes the vertex at the center and the sum is over the other vertices. The quantum processor is initialized with the center qubit in the $1$ state and the remaining qubits are in the $0$ state. In the staggered fermion lattice regularization, sites are alternatively identified with matter and anti-matter degrees of freedom so this state should correspond to the trivial vacuum. By evolving with the Hamiltonian in Eq.~\eqref{eq:effQCDV}, it should be possible to observe matter anti-matter fluctuations. Note that with this initial state, a single Trotter step can be performed without having to implement CNOT gates from the Jordan-Wigner strings. A single Trotter step was implemented on IBM {\tt Perth} with the size of the time step being varied to sample different times. Due to the connectivity of the hardware, this circuit required 28 CNOT gates. Fig.~\ref{fig:Perth3D} shows the results of performing a single Trotter step for $g=2$ on IBM {\tt Perth}. For small times, the quantum simulation is able to describe the evolution of the system accurately, however beyond $t=1$, the error in the single Trotter step used is large and limits the accuracy of the quantum simulation.

While the Jordan-Wigner encoding is efficient in the number of qubits used, the Hamiltonian generated has long range interactions which are necessary to preserve the anti-commutation relation of the fermions. Scaling these calculations to a larger lattice will require making use of a more efficient fermion encoding. For example, the Bravyi-Kitaev superfast encoding can be used to map fermions onto qubits~\cite{Setia_2019}. In this encoding, a qubit is associated with each link on the lattice and represents the parity of the number of fermions on the link. The length of the strings of Pauli $\hat{Z}$ operators for an operator on a link extends only to neighboring links. For a large lattice, this will limit the circuit depth necessary to perform time evolution and potentially allow for larger calculations to be performed.

\section{Discussion}
\noindent
In this work, the SRG has been used to derive improved Hamiltonians that mitigate the effects of gauge field truncation. It was demonstrated in 1+1D that the improved Hamiltonians derived this way outperform those derived through the strong coupling expansion for small systems. Tensor network calculations were performed to demonstrate that these improved Hamiltonians perform well as the system size is increased. These techniques were also applied to 3+1D giving an improved Hamiltonian capable of describing two flavour QCD on the lattice. Real time dynamics on small systems were simulated on IBM's {\tt Perth} quantum processor.

Previous strategies for quantum simulation of lattice gauge theories improved accuracy by increasing the truncation of the gauge field. This comes at the cost of needing more qubits to represent the system and a more complicated circuit to implement the time evolution. The improved Hamiltonians introduced in this work are capable of improving accuracy only at the cost of requiring more complicated circuits to simulate.

Improved Hamiltonians have been derived for a single flavor of staggered fermions coupled to SU(3) gauge fields truncated at low electric field. This has enabled quantum simulation of systems that would otherwise be out of reach of current quantum hardware. The same approach introduced here can be used to derive improved Hamiltonians for larger electric field truncations and with more flavors of fermions. Future work will extend these methods to higher spatial dimensions with larger operator truncations where the plaquette terms will modify the SRG flow. This will enable quantum simulations of lattice gauge theories in multiple dimensions to be performed in the near term.

\begin{acknowledgments}
The authors would like to acknowledge useful conversations about SRG with Zhiyao Li on a related project. We would also like to thank Marc Illa and Roland Farrell for feedback in preparing this manuscript. The authors would also like to acknowledge many useful conversations with Martin Savage, Francesco Turro, Xiaojun Yao and Niklas Mueller. The material presented here was funded by U.S. Department of Energy, Office of Science, Office of Nuclear Physics, Inqubator for Quantum Simulation (IQuS)\footnote{\url{https://iqus.uw.edu}} under Award Number DOE (NP) Award DE-SC0020970 via the program on Quantum Horizons: QIS Research and Innovation for Nuclear Science. This work was enabled, in part, by the use of advanced computational, storage and networking infrastructure provided by the Hyak supercomputer system at the University of Washington\footnote{\url{https://itconnect.uw.edu/research/hpc}}. We acknowledge the use of IBM Quantum services for this work. The views expressed are those of the authors, and do not reflect the official policy or position of IBM or the IBM Quantum team.
\end{acknowledgments}

\bibliography{ref}



\appendix
\section{Schrieffer-Wolff Perturbation Theory}
\label{appendix:SWPT}
The improved Hamiltonians derived in this work are based on performing a unitary transformation before truncating the electric field to reduce the coupling to the states being removed by the truncation. This can be done perturbatively through the use of Schrieffer-Wolf perturbation theory (SWPT). In this section, the application of SWPT to the Hamiltonian in Eq.~\eqref{eq:QCD1DNoGauge} with $m=0$ will be demonstrated. The Hamiltonian for lattice gauge theories in 1D we wish to simulate takes the form
\begin{equation}
    \hat{H} = \hat{H}_E + \hat{H}_{D} + \hat{V} \ \ \ ,
\end{equation}
where $\hat{H}_E$ is the electric Hamiltonian, $\hat{V}$ couples the low energy subspace to the high energy subspace and $\hat{H}_{D}$ describes dynamics in the high energy Hilbert space. Note that the kinetic term of Eq.~\eqref{eq:QCD1DNoGauge} is equal to $\hat{H}_{D} + \hat{V}$. For the zero electric field truncation, $\hat{V}$ is the piece of the kinetic term that corresponds to a baryon on a site ejecting a quark to a neighboring site and $\hat{H}_D$ is the piece of the kinetic term that describes a quark propagating freely between sites. SWPT systematically generates a unitary, $e^{\hat{S}}$ that decouples the selected low energy subspace. For lattice gauge theories, we will be decoupling the electric vacuum and states with low energy relative to the electric Hamiltonian. To leading order we have
\begin{align}
    e^{\hat{S}_1} \hat{H} e^{-\hat{S}_1} & = \hat{H}_E + [\hat{S}_1,\hat{H}_E + \hat{H}_{D}] + \hat{H}_{D} + \hat{V}  + [\hat{S}_1,\hat{V}] \nonumber \\
    & + \frac{1}{2}[\hat{S}_1,[\hat{S}_1,\hat{H}_E + \hat{H}_{D}] + \mathcal{O}(\hat{V}^3)\ \ \ .
\end{align}
The leading order coupling between the low and high energy subspace comes from $\hat{V}$ and be cancelled at leading order by choosing $\hat{S}_1$ such that $ [\hat{S}_1,\hat{H}_E + \hat{H}_{D}]=-\hat{V}$. Explicitly, the matrix elements of $\hat{S}_1$ are 
\begin{equation}
    (S_1)_{ab} = \frac{1}{E_a - E_b} V_{ab} \ \ \ ,
\end{equation}
where the indices label eigenstates of $\hat{H}_E + \hat{H}_{D}$ with eigenvalues $E_a$. To leading order, the effective Hamiltonian is
\begin{equation}
    \hat{H}_{eff}^1 = \hat{H}_E + \frac{1}{2}[\hat{S}_1,\hat{V}] \ \ \ ,
\end{equation}
and provided that the low energy subspace has an electric energy of 0, the commutator is equal to
\begin{equation}
    \frac{1}{2}[\hat{S}_1,\hat{V}] = -\hat{V} \frac{1}{\hat{H}_E + \hat{H}_{D}} \hat{V}  = -\sum_n \hat{V} \left(-\hat{H}_E^{-1} \hat{H}_{D}\right)^n \frac{1}{\hat{H}_E} \hat{V} \ \ \ . 
\end{equation}
Therefore to $\mathcal{O}(H_E^{-2})$, the effective Hamiltonian is given by
\begin{equation}
    \hat{H}_{eff}^1 = \hat{H}_E - \hat{V} \frac{1}{\hat{H}_E} \hat{V} + \hat{V} \frac{1}{\hat{H}_E} \hat{H}_{D} \frac{1}{\hat{H}_E} \hat{V} + \mathcal{O}(\hat{H}_E^{-3})\ \ \ .
    \label{eq:SWLeading}
\end{equation}

To derive Eq.~\eqref{eq:HeffSW} from Eq.~\ref{eq:SWLeading}, it will be helpful to work in the gauge invariant subspace of the theory. The gauge invariant states of the two staggered site Hamiltonian are
\begin{align}
    \ket{\Omega} & = \hat{\psi}^\dagger_{1,b} \hat{\psi}^\dagger_{1,g} \hat{\psi}^\dagger_{1,r}\ket{0} \nonumber \\
    \ket{M} & = \frac{1}{\sqrt{3}} \sum_c \hat{\psi}^\dagger_{2,c} \hat{\psi}_{1,c}\ket{\Omega} \nonumber \\
    \ket{T} & =  \frac{1}{2\sqrt{3}} \sum_{c'} \hat{\psi}^\dagger_{2,c'} \hat{\psi}_{1,c'} \sum_c \hat{\psi}^\dagger_{2,c} \hat{\psi}_{1,c}\ket{\Omega} \nonumber \\
    \ket{B} & = \hat{\psi}^\dagger_{2,b} \hat{\psi}^\dagger_{2,g} \hat{\psi}^\dagger_{2,r}\ket{\Omega}\nonumber \\
    \ket{\Bar{B}} & = \hat{\psi}_{1,b} \hat{\psi}_{1,g} \hat{\psi}_{1,r}\ket{\Omega} \nonumber \\
    \ket{B \Bar{B}} & = \hat{\psi}_{1,b} \hat{\psi}_{1,g} \hat{\psi}_{1,r} \hat{\psi}^\dagger_{2,b} \hat{\psi}^\dagger_{2,g} \hat{\psi}^\dagger_{2,r} \ket{\Omega} \ \ \ ,
\end{align}
where $\ket{0}$ is the state with all fermion modes unoccupied, $\hat{\psi}_{x,c}$ is the quark field operator on site $x$ with color $c$, $\ket{\Omega}$ is the electric vacuum, $\ket{M}$ is a meson state in the strong coupling limit with massive quarks, $\ket{T}$ is a tetra-quark state, $\ket{B}$ is a state with a single baryon, $\ket{\Bar{B}}$ is a state with a single anti-baryon and $\ket{B \Bar{B}}$ is a state with a baryon anti-baryon pair. When these basis states are enumerated in the order $\{\ket{\Bar{B}},\ket{\Omega},\ket{B\Bar{B}},\ket{B},\ket{M},\ket{T}\}$, the electric energy operator written as a matrix is
\begin{equation}
    \hat{H}_E = \begin{pmatrix}
        0 & 0 & 0 & 0 & 0 & 0 \\
        0 & 0 & 0 & 0 & 0 & 0 \\
        0 & 0 & 0 & 0 & 0 & 0 \\
        0 & 0 & 0 & 0 & 0 & 0 \\
        0 & 0 & 0 & 0 & \frac{2}{3}g^2 & 0 \\
        0 & 0 & 0 & 0 & 0 & \frac{2}{3}g^2 
    \end{pmatrix}
\end{equation}
and the kinetic term is
\begin{equation}
    \hat{H}_K = \begin{pmatrix}
        0 & 0 & 0 & 0 & 0 & 0 \\
        0 & 0 & 0 & 0 & \frac{\sqrt{3}}{2} & 0 \\
        0 & 0 & 0 & 0 & 0 & \frac{\sqrt{3}}{2} \\
        0 & 0 & 0 & 0 & 0 & 0 \\
        0 & \frac{\sqrt{3}}{2} & 0 & 0 & 0 & 1 \\
        0 & 0 & \frac{\sqrt{3}}{2} & 0 & 1 & 0 
    \end{pmatrix} \ \ \ .
\end{equation}
Note that this ordering of basis states has been chosen so that the encoding onto qubits corresponds to the upper left $4\times4$ corner of this matrix. The decomposition of $\hat{H}_K$ into $\hat{V} +\hat{H}_D$ is given by
\begin{equation}
    \hat{V} = \begin{pmatrix}
        0 & 0 & 0 & 0 & 0 & 0 \\
        0 & 0 & 0 & 0 & \frac{\sqrt{3}}{2} & 0 \\
        0 & 0 & 0 & 0 & 0 & \frac{\sqrt{3}}{2} \\
        0 & 0 & 0 & 0 & 0 & 0 \\
        0 & \frac{\sqrt{3}}{2} & 0 & 0 & 0 & 0 \\
        0 & 0 & \frac{\sqrt{3}}{2} & 0 & 0 & 0 
    \end{pmatrix} \ \ \ ,
\end{equation}
and
\begin{equation}
    \hat{H}_D = \begin{pmatrix}
        0 & 0 & 0 & 0 & 0 & 0 \\
        0 & 0 & 0 & 0 & 0 & 0 \\
        0 & 0 & 0 & 0 & 0 & 0 \\
        0 & 0 & 0 & 0 & 0 & 0 \\
        0 & 0 & 0 & 0 & 0 & 1 \\
        0 & 0 & 0 & 0 & 1 & 0 
    \end{pmatrix} \ \ \ .
\end{equation}
Truncating to the basis states $\{\ket{\Bar{B}},\ket{\Omega},\ket{B\Bar{B}},\ket{B}\}$, the leading term in the effective Hamiltonian is given by
\begin{equation}
    - \hat{V} \frac{1}{\hat{H}_E} \hat{V} = \begin{pmatrix}
        0 & 0 & 0 & 0 \\
        0 & -\frac{9}{8g^2} & 0 & 0 \\
        0 & 0 & -\frac{9}{8g^2} & 0 \\
        0 & 0 & 0 & 0 
    \end{pmatrix} \ \ \ ,
    \label{eq:SWterm1}
\end{equation}
and after dropping a term proportional to the identity matrix this is equivalent to $\frac{9}{16 g^2} \hat{Z}_1 \hat{Z}_2$. Similarly, the subleading term is
\begin{equation}
    \hat{V} \frac{1}{\hat{H}_E} \hat{H}_{D} \frac{1}{\hat{H}_E} \hat{V} = \begin{pmatrix}
        0 & 0 & 0 & 0 \\
        0 & 0 & \frac{27}{16g^4} & 0 \\
        0 & \frac{27}{16g^4} & 0 & 0 \\
        0 & 0 & 0 & 0 
    \end{pmatrix} \ \ \ ,
    \label{eq:SWterm2}
\end{equation}
and written in terms of Pauli matrices this is $\frac{27}{32g^4}\left(\hat{X}_1 \hat{X}_2 + \hat{Y}_1 \hat{Y}_2 \right)$. Adding together the expressions in Eq.~\ref{eq:SWterm1} and Eq.~\ref{eq:SWterm2} yields the improved Hamiltonian in Eq~\ref{eq:HeffSW}. Techniques for performing this expansion to higher orders can be found in Ref.~\cite{bravyi2011schrieffer}.

\end{document}